\documentclass[english]{article}
\usepackage[T1]{fontenc}
\usepackage[latin9]{inputenc}
\usepackage{amsbsy}
\usepackage{amssymb}
\usepackage{graphicx}
\usepackage{esint}
\usepackage{babel}
\begin{document}
\title{3D Topological Quantum Computing}
\author{Torsten Asselmeyer-Maluga\\German Aerospace Center (DLR), Rosa-Luxemburg-Str. 2\\
10178 Berlin, Germany\\
torsten.asselmeyer-maluga@dlr.de}
\maketitle
\begin{abstract}
In this paper we will present some ideas to use 3D topology for quantum
computing extending ideas from a previous paper. Topological quantum
computing used \textquotedblleft knotted\textquotedblright{} quantum
states of topological phases of matter, called anyons. But anyons
are connected with surface topology. But surfaces have (usually) abelian
fundamental groups and therefore one needs non-abelian anyons to use
it for quantum computing. But usual materials are 3D objects which
can admit more complicated topologies. Here, complements of knots
do play a prominent role and are in principle the main parts to understand
3-manifold topology. For that purpose, we will construct a quantum
system on the complements of a knot in the 3-sphere (see arXiv:2102.04452
for previous work). The whole system is designed as knotted superconductor
where every crossing is a Josephson junction and the qubit is realized
as flux qubit. We discuss the properties of this systems in particular
the fluxion quantization by using the A-polynomial of the knot. Furthermore
we showed that 2-qubit operations can be realized by linked (knotted)
superconductors again coupled via a Josephson junction.
\end{abstract}

\section{Introduction}

Quantum computing exploits quantum-mechanical phenomena such as superposition
and entanglement to perform operations on data, which in many cases,
are infeasible to do efficiently on classical computers. Topological
quantum computing seeks to implement a more resilient qubit by utilizing
non-Abelian forms of matter like non-abelian anyons to store quantum
information. Then, operations (what we may think of as quantum gates)
are performed upon these qubits through \textquotedblleft braiding\textquotedblright{}
the worldlines of the anyons. We refer to the book \cite{topoQuantComp}
for an introduction of these ideas.

In a previous paper \cite{AsselmeyerMaluga2021a} we described first
ideas to use 3-manifold topology for topological quantum computing.
There, we discussed the knot complement as the main part of a 3-manifold.
Main topological invariant of a knot complement is the fundamental
group \cite{GordonLuecke1989}. In the previous paper we discussed
the representation of the fundamental group into $SU(2)$ to produce
the 1-qubit operations. Furthermore we argued that linking is able
to produce 2-qubit operations showing the universality of the approach.
The operations are realized by the consideration of the Berry phase.
In contrast, in this paper we will realize 1-qubit operations of the
knot group via knotted superconductors. Here, the crossings are given
by Josephson junctions to mimic the knot. Furthermore, the 2-qubit
operations is again realized by the linking of two knotted superconductors
via a Josephson element, in agreement with \cite{AsselmeyerMaluga2021a}.

In the next section we will introduce the concept of a fundamental
group and manifold. Furthermore we explain the importance of knot
complements. In section \ref{sec:Knot-groups} we explain the representation
of knot groups partly from \cite{AsselmeyerMaluga2021a} to make the
paper self-contained. Then in section \ref{sec:Knotted-superconductors}
we introduce the model of a knotted superconductor with flux qubit.
Here we will discuss the fluxion quantization (using the A-polynomial)
and the influence of the Josephson junction. Main result is the realization
of the 2-qubit operation by linking.

\section{Some preliminaries and motivation: 3-manifolds and knot complements}

In this paper we need two concepts, the fundamental group and the
manifold, which will be introduced now. We assume that the reader
is familiar with the definition of a topological space. Let $X,Y$
be topological spaces. First consider the definition of homotopy as
applied to a pair of maps, $f,g:X\to Y.$ 
\begin{itemize}
\item Let $f,g:X\to Y$ be continuous functions. $f$ and $g$ are homotopic
to each other, denoted by $f\simeq g$ if there is a continuous function
$F:X\times[0,1]\to Y$ with $F(x,0)=f(x)$ and $F(x,1)=g(x)$ for
all $x\in X$.
\item The function $F$ provides a deformation of one map into the other.
Clearly, this relation is an equivalence relation. The equivalence
class of homotopic maps between $X$ and $Y$ will be denoted by 
\[
[X,Y]=\{f:X\to Y\mbox{ continuous }\}/\simeq\quad.
\]
This relation leads to the notion of homotopy-equivalence of spaces. 
\item Two topological spaces $X$ and $Y$ are homotopy-equivalent, if there
are two smooth maps $f:X\to Y$ and $g:Y\to X$ so that 
\[
f\circ g\simeq Id_{Y}\qquad g\circ f\simeq Id_{X}
\]
where $Id_{X}$ and $Id_{Y}$ are the identity maps on $X$ and $Y$,
respectively. 
\end{itemize}
In general define 
\[
\pi_{n}(X,x_{0})=[(S^{n},s_{0}),(X,x_{0})]\quad,
\]
the homotopy equivalence class of maps of the pointed sphere into
the pointed space. Since we have used the word ``group'' to refer
to them we must define a combining operation. 

Here we need the $n=1$ case, $\pi_{1},$ the fundamental group. This
is the loop space, modulo smooth deformations or contractions. There
are several ways to define the group combining operation. We choose
one which is easily extendable from $n=1$ to the general case. Let
$S^{1}\vee S^{1}$ be the one-point union defined by 
\[
S^{1}\vee S^{1}=S^{1}\times\{s_{1}\}\cup_{s_{0}\equiv s_{1}}\{s_{0}\}\times S^{1}\subset S^{1}\times S^{1}\ni(s_{0},s_{1}).
\]
Now define the product $\gamma_{1}\star\gamma_{2}:S^{1}\to X$ of
the maps $\gamma_{1},\gamma_{2}:S^{1}\to X$ to $\gamma_{12}:S^{1}\to X$
as defined geometrically by the process to identify two opposite points
on the circle $S^{1}$, with naturally defined map. We then combine
this map with the maps $\gamma_{1},\ \gamma_{2}$ to define the product,
$\gamma_{1}\star\gamma_{2}$. This provides a group product structure
which will in general not be abelian, since there is no map homotopic
to the identity which switches the upper and lower circles in this
diagram. The proof of the associativity of this product can be found
for instance in \cite{Bre:93} Proposition 14.16. These formal definitions
have a simple interpretation:
\begin{itemize}
\item elements of the fundamental group $\pi_{1}(X)$ are closed non-contractible
curves (the unit element $e$ is the contractible curve);
\item the group operation is the concatenation of closed curves up to homotopy
(to guarantee associativity);
\item the inverse group element is a closed curve with opposite orientation;
\item the fundamental group $\pi_{1}(X)$ is a topological invariant, i.e.
homeomorphic spaces $X,Y$ have isomorphic fundamental groups;
\item the space $X$ is (usually) path-connected so that the choice of the
point $x_{0}$ is arbitrary.
\end{itemize}
Clearly, every closed curve in $\mathbb{R}^{2},\mathbb{R}^{3}$ is
contractible, therefore $\pi_{1}(\mathbb{R}^{2})=0=\pi_{1}(\mathbb{R}^{3})$.
The group is nontrivial for the circle $\pi_{1}(S^{1})$. Obviously,
a curve $a$ going around the circle is a closed curve which cannot
be contract (to a point). The same is true for the concatenation $a^{2}$
of two closed curves, three curves $a^{3}$ etc. Therefore a closed
curve in $S^{1}$ is characterized by the winding number of the closed
curve or $\pi_{1}(S^{1})=\mathbb{Z}$, see \cite{Bre:93} for more
details. In general the fundamental group consists of sequences of
generators (the alphabet) restricted by relations (the grammar). For
$\pi_{1}(S^{1})$, we have one generator $a$ but no relations, i.e.
\[
\pi_{1}(S^{1})=\langle a|\,\emptyset\rangle=\mathbb{Z}
\]
A similar argumentation can be used for $S^{1}\vee S^{1}$, the one-point
union of two circles. The closed curve in the first circle is generator
$a$ and in the second circle it is $b$. There is no relation, i.e.
\[
\pi_{1}(S^{1}\vee S^{1})=\langle a,b|\,\emptyset\rangle=\mathbb{Z}*\mathbb{Z}
\]
the group is non-abelian (the sequences $ab$ and $ba$ are different).
The second ingredient of our work is the concept of a manifold defined
by:
\begin{itemize}
\item Let $M$ be a Hausdorff topological space covered by a (countable)
family of open sets, ${\mathcal{U}}$, together with homeomorphisms,
$\phi_{U}:{\mathcal{U}}\ni U\rightarrow U_{R},$ where $U_{R}$ is
an open set of ${\mathbb{R}}^{n}.$ This defines $M$ as a topological
manifold. For smoothness we require that, where defined, $\phi_{U}\cdot\phi_{V}^{-1}$
is smooth in ${\mathbb{R}}^{n},$ in the standard multi-variable calculus
sense. The family ${\mathcal{A}}=\{{\mathcal{U}},\phi_{U}\}$ is called
an atlas or a differentiable structure. Obviously, ${\mathcal{A}}$
is not unique. Two atlases are said to be compatible if their union
is also an atlas. From this comes the notion of a maximal atlas. Finally,
the pair $(M,{\mathcal{A}})$, with ${\mathcal{A}}$ maximal, defines
a smooth manifold of dimension $n$. 
\end{itemize}
Now we will concentrate on two- and three-dimensional manifolds, 2-manifold
and 3-manifold for short. For 2-manifolds, the basic elements are
the 2-sphere $S^{2}$, the torus $T^{2}$ or the Klein bottle $\mathbb{R}P^{2}$.
Then one gets for the classification of 2-manifolds:
\begin{itemize}
\item Every compact, closed, oriented 2-manifold $S_{g}$ is homeomorphic
to either $S^{2}$ ($\pi_{1}(S^{2})=0$) or to the connected sum 
\[
S_{g}=\underbrace{T^{2}\#T^{2}\#\ldots\#T^{2}}_{g}\qquad\pi_{1}(S_{g})=\underbrace{\mathbb{Z}\oplus\cdots\oplus\mathbb{Z}}_{2g}
\]
of $T^{2}$ for a fixed genus $g$. Every compact, closed, non-oriented
2-manifold is homeomorphic to the connected sum 
\[
\underbrace{{\mathbb{R}}P^{2}\#{\mathbb{R}}P^{2}\#\ldots\#{\mathbb{R}}P^{2}}_{g}\qquad\pi_{1}(\tilde{S}_{g})=\langle a_{1},\ldots,a_{g}\,|\,a_{1}^{2}\cdots a_{g}^{2}=e\rangle
\]
of ${\mathbb{R}}P^{2}$ for a fixed genus $g$.
\item Every compact 2-manifold with boundary can be obtained from one of
these cases by cutting out the specific number of disks $D^{2}$ from
one of the connected sums. 
\end{itemize}
The connected sum operation $\#$ is defined by: Let $M,N$ be two
$n$-manifolds with boundaries $\partial M,\partial N$. The connected
sum $M\#N$ is the procedure of cutting out a disk $D^{n}$ from the
interior $int(M)\setminus D^{n}$ and $int(N)\setminus D^{n}$ with
the boundaries $S^{n-1}\sqcup\partial M$ and $S^{n-1}\sqcup\partial N$,
respectively, and gluing them together along the common boundary component
$S^{n-1}$. This operation is important for 3-manifolds too. But the
classification of 3-manifolds is more complex. Following {Thurston\lq s
idea, one needs eight pieces which are arranged by using two sums
(sum along a torus and connected sum). We don\lq t want} go into the details
and refer to \cite{Scott1983,Thu:97} for a description of 3-manifolds
(the conjecture of Thurston was proved by Perelman \cite{Per:02,Per:03.1,Per:03.2}
in 2003). The following facts are a consequence of this classification:
\begin{itemize}
\item 3-manifolds are mainly classified by the fundamental group (and the
Reidemeister torsion for lens spaces),
\item the fundamental groups of 3-manifolds can be non-abelian which is
impossible for oriented 2-manifolds
\item the simplest pieces of 3-manifolds are mainly given by the complement
of a knot or link.
\end{itemize}
Even the last point is the main motivation of this paper. In contrast
to topological quantum computing with anyons, we cannot directly use
3-manifolds (as submanifolds) like surfaces in the fractional Quantum
Hall effect. Surfaces (or 2-manifolds) embed into a 3-dimensional
space like $\mathbb{R}^{3}$ but 3-manifolds require a 5-dimensional
space like $\mathbb{R}^{5}$ as an embedding space. Therefore, we
cannot directly use 3-manifolds. However, as we argued above, there
is a group-theoretical substitute for a 3-manifolds, the fundamental
group of a knot complement also known as knot group which will be
introduced now.

A knot in mathematics is the embedding $K:S^{1}\to S^{3}$ of a circle
into the 3-sphere $S^{3}$ (or in $\mathbb{R}^{3}$), i.e. a closed
knotted curve $K(S^{1})$ (or $K$ for short). To form the knot complement,
we have to consider a thick knot $K\times D^{2}$ (knotted solid torus).
Then the knot complement is defined by
\[
C(K)=S^{3}\setminus(K\times D^{2})
\]
and the knot group $\pi_{1}(C(K))$ is the fundamental group of the
knot complement. The knot complement $C(K)$ is a 3-manifold with
boundary $\partial C(K)=T^{2}$. It was shown that prime knots are
divided into two classes: hyperbolic knots ($C(K)$ admits a hyperbolic
structure) and non-hyperbolic knots ($C(K)$ admits one of the other
seven geometric structures). An embedding of disjoints circles into
$S^{3}$ is called a link $L.$ Then, $C(L)$ is the link complement.
If we speak about 3-manifolds then we have to consider $C(K)$ as
one of the basic pieces. Furthermore, there is the Gordon--Luecke
theorem: if two knot complements are homeomorphic, then the knots
are equivalent (see in \cite{GordonLuecke1989} for the statement
of the exact theorem). Interestingly, knot complements of prime knots
are determined by its fundamental group.

\section{Knot groups and quantum computing representations\label{sec:Knot-groups}}

Any knot can be represented by a projection on the plane with no multiple
points which are more than double. As an example let us consider the
simplest knot, the trefoil knot $3_{1}$ (knot with three crossings).
\begin{figure}[t]
\centerline{\includegraphics[scale=0.1]{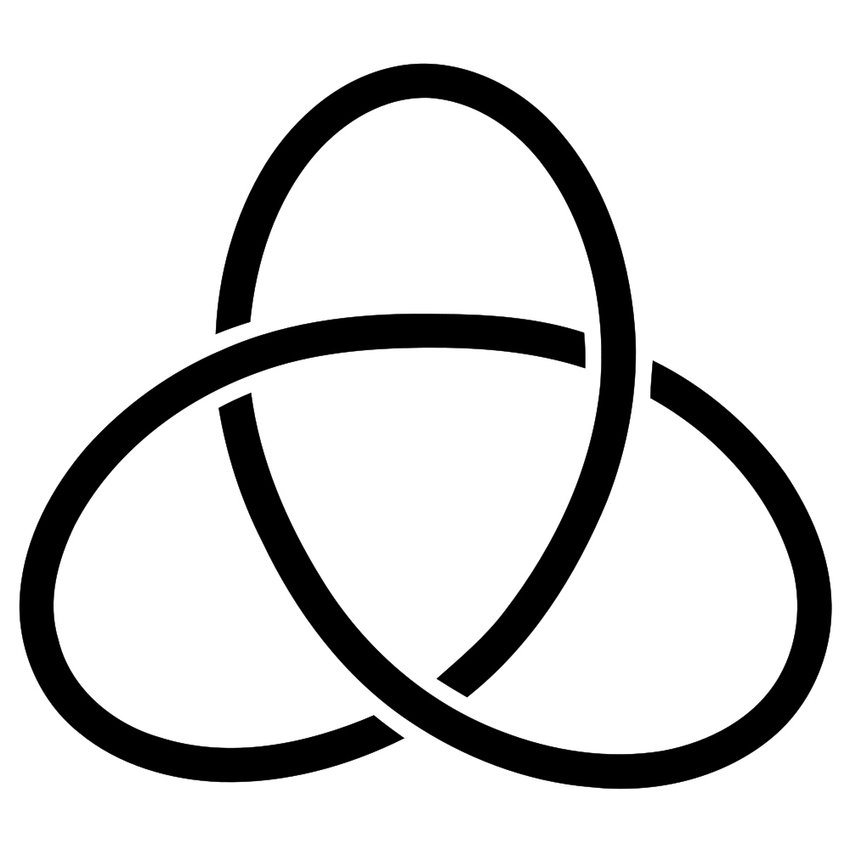}}
\vspace*{8pt}
\caption{the simplest knot, trefoil knot $3_{1}$ \label{fig:trefoil-knot}}
\end{figure}
 The plane projection of the trefoil is shown in Fig. \ref{fig:trefoil-knot}.
This projection can be divided into three arcs, around each arc we
have a closed curve as generator of $\pi_{1}(C(3_{1}))$ denoted by
$a,b,c$ (see Fig. \ref{fig:three-generators-knot-group-trefoil}).
\begin{figure}[t]
\centerline{\includegraphics[scale=0.25]{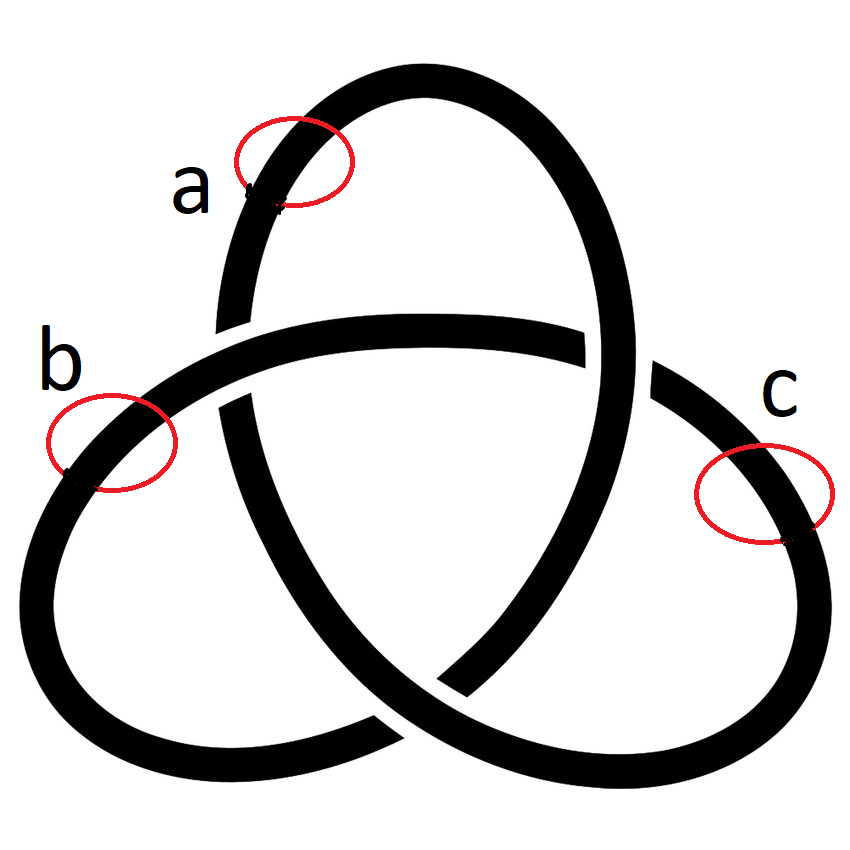}}
\vspace*{8pt}
\caption{the three generators $a,b,c$ of the knot group of the trefoil knot
$3_{1}$\label{fig:three-generators-knot-group-trefoil}}

\end{figure}
Now each crossing gives a relation between the corresponding generators:
$c=a^{-1}ba,b=c^{-1}ac,a=b^{-1}cb$, i.e. we obtain the knot group
\[
\pi_{1}(C(3_{1}))=\langle a,b,c\,|\,c=a^{-1}ba,b=c^{-1}ac,a=b^{-1}cb\rangle
\]
Then we substitute the expression $c=a^{-1}ba$ into the other relations
to get a representation of the knot with two generators and one relation.
From relation $a=b^{-1}cb$ we will obtain $a=b^{-1}(a^{-1}ba)b$
or $bab=aba$ and the other relation $b=c^{-1}ac$ gives nothing new.
Finally we will get the result \cite{Rol:76,KauffmanKnotsPhysics}
\[
\pi_{1}(C(3_{1}))=\langle a,b\,|\,bab=aba\rangle
\]
But this group is also well-known, it is the braid group $B_{3}$
of three strands. Now we will get in touch with quantum computing.
The main idea is the interpretation of the braid group $B_{3}$ as
operations (gates) on qubits. From the mathematical point of view,
we have to consider the representation of $B_{3}$ into $SU(2)$,
i.e. a homomorphism
\[
\phi:B_{3}\to SU(2)
\]
mapping sequences of generators (called words) into matrices as elements
of $SU(2)$. At first we note that a matrix in $SU(2)$ has the form
\[
M=\left(\begin{array}{cc}
z & w\\
-\bar{w} & \bar{z}
\end{array}\right)\qquad|z|^{2}+|w|^{2}=1
\]
where $z$ and $w$ are complex numbers. Now we choose a well-known
basis of $SU(2)$:
\[
\boldsymbol{1}=\left(\begin{array}{cc}
1 & 0\\
0 & 1
\end{array}\right)\quad\boldsymbol{i}=\left(\begin{array}{cc}
i & 0\\
0 & -i
\end{array}\right)\quad\boldsymbol{j}=\left(\begin{array}{cc}
0 & 1\\
-1 & 0
\end{array}\right)\quad\boldsymbol{k}=\left(\begin{array}{cc}
0 & i\\
i & 0
\end{array}\right)
\]
so that 
\[
M=a\boldsymbol{1}+b\boldsymbol{i}+c\boldsymbol{j}+d\boldsymbol{k}
\]
with $a^{2}+b^{2}+c^{2}+d^{2}=1$ (and $z=a+bi,w=c+di$). The algebra
of $1,\boldsymbol{i},\boldsymbol{j},\boldsymbol{k}$ are known as
quaternions. Among all representations, there is the simplest example
\[
g=e^{7\pi\boldsymbol{i}/10},\:f=\boldsymbol{i}\tau+\boldsymbol{k}\sqrt{\tau},\:h=fgf^{-1}
\]
where $\tau^{2}+\tau=1$ and we have the matrix representation 
\[
g=\left(\begin{array}{cc}
e^{i7\pi/10} & 0\\
0 & e^{-i7\pi/10}
\end{array}\right)\qquad f=\left(\begin{array}{cc}
i\tau & i\sqrt{\tau}\\
i\sqrt{\tau} & -i\tau
\end{array}\right)
\]
Then $g,h$ satisfy $ghg=hgh$ the relation of $B_{3}$. This representation
is known as the Fibonacci representation of $B_{3}$ to $SU(2)$.
The Fibonacci representation is dense in $SU(2)$, see \cite{KauffmanLomonaco2008,topoQuantComp}. {The Fibonacci representation is usually used in anyonic quantum computing. It denotes a special representation of the braid group into $SU(2)$. In case of the trefoil knot, the knot group is the braid group $B_3$ so that the representation of the knot group agrees with the representation in anyonic quantum computing. The 3-manifold associated to the trefoil knot is the Poincare sphere.}

\begin{figure}[t]
\centerline{\includegraphics[scale=0.5]{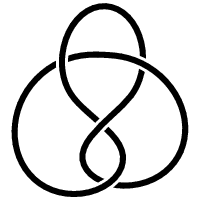}}
\vspace*{8pt}
\caption{figure-8 knot $4_{1}$. \label{fig:figure-8-knot}}
\end{figure}

However, there are more complicated knots. The complexity of knots
is measured by the number of crossings. There is only one knot with
three crossings (trefoil) and with four crossings (figure-8). For
the figure-8 knot $4_{1}$ (see Figure \ref{fig:figure-8-knot}),
the knot group is given by 
\[
\pi_{1}\left(C(4_{1})\right)=\langle a,b\,|\:bab^{-1}ab=aba^{-1}ba\rangle
\]
admitting a representation $\phi$ into $SU(2)$, see \cite{KirKla:90}.
Here, we remark that the figure-8 knot is part of a large class, the
so-called hyperbolic knots. Hyperbolic knots are characterized by
the property that the knot complement admits a hyperbolic geometry.
Hyperbolic knot complements have special properties, in particular
topology and geometry are connected in a special way. Central property
is the so-called Mostow-Prasad rigidity \cite{Mos:68,Prasad:1973}:
every deformation of the space is an isometry, or geometric properties
like volume or curvature are topological invariants. In particular,
the hyperbolic structure can be used to get new invariants (but only
for hyperbolic knots). A well-known invariant is the A-polynomial
\cite{APolynomial1994,APolynomial1998} which will be used in the
next section.

\section{Knotted superconductors and knot groups\label{sec:Knotted-superconductors}}

{In this section we will present first ideas to realize the knot complement
by using a superconducting ring which is knotted. \\
The usage of superconductors for quantum computing is divided into three possible realizations: charge qubit, flux qubit and phase qubit but also many hybridizations exist like Fluxonium \cite{Fluxonium2009}, Transmon \cite{Transmon2009}, Xmon \cite{Xmon2013} and Quantronium \cite{Quantronium2007}. The qubit implementation as the logical quantum states 
$|0\rangle,|1\rangle$ is realized by the mapping to the different states of the system. Therefore we have to deal with the states in the knotted superconductor. For a superconductor, there is a single wave function of the condensate. For the qubit realization, we have to consider a superposition of two wave functions in different energy states.
The knotted superconductor has Josephson junctions at the over and under crossings. Let $\psi_1$ and $\psi_2$ be the wave functions at the over or under crossing. By using the Ginsburg-Landau theory we will get the current
$$
j=\frac{e\hbar V}{m}|\psi_1|^2\sin(\Phi_{12})
$$
where $V$ is the potential difference and $\Phi_{12}$ is the phase difference between the two wave functions. If the knotted superconductor consists of a single state, say $|0\rangle$, then the Josephson junctions at the over or under crossings have no effect because there is no potential difference. 
In case of the two state $|0\rangle$ and $|1\rangle$, there is a energy difference and one gets 
a coupling between the two states $|0\rangle,|1\rangle$ which is given by
\[
i\hbar\frac{d}{dt}\left(\begin{array}{c}
|0\rangle\\
|1\rangle
\end{array}\right)=\left(\begin{array}{cc}
E_1 & Ve^{i\phi}\\
Ve^{-i\phi} & E_2
\end{array}\right)\left(\begin{array}{c}
|0\rangle\\
|1\rangle
\end{array}\right)=H\left(\begin{array}{c}
|0\rangle\\
|1\rangle
\end{array}\right)
\]
where $V$ depends on the energy difference $E_1-E_2$ and the coupling between the states.
Then we get the Hamiltonian
\[
H_{loop}= E_1\frac{\boldsymbol{1}+\sigma_z}{2}+E_2\frac{\boldsymbol{1}-\sigma_z}{2}+V\cdot\cos\phi\cdot\sigma_{x}+V\cdot\sin\phi\cdot\sigma_{y}
\]
acting on the single state in terms of Pauli matrices (as generators
of the Lie algebra of $SU(2)$). The same Hamiltonian also works for Josephson
junctions which are placed at the generators $a,b,c$ of the knot group (see Fig. \ref{fig:three-generators-knot-group-trefoil}).
In the previous section we described the representation of the generators abstractly.
Here, the energy and coupling determines the representation. Now let us choose the
Josephson junction for the generator $a$. Furthermore we trim the energy levels via the
junction so that $|0\rangle$ has energy $E$ and $|1\rangle$ has energy $-E$. 
For simplicity we neglect the phase shift $\phi=0$. Then we obtain the Hamiltonian 
\[
H_{1}= E\cdot\sigma_{z}+V\cdot\sigma_{x}
\] 
and for small couplings one gets the generator $g$ of the Fibonacci representation 
(see the previous section). If the coupling is stronger, then we get in principle the other generator $f$ of the Fibonacci representation (by a suitable choice of the energy). \\
In a previous paper \cite{AsselmeyerMaluga2021a}
we constructed the 1-qubit operations from the knot complement. Furthermore
we argued that the linking of two knots is needed to generate the
2-qubit operations. The simplest link is the Hopf link (denoted as
$L2a1$, see Figure \ref{fig:Hopf-link}), the linking of two unknotted
curves. 
\begin{figure}[t]
\centerline{\includegraphics[scale=0.4]{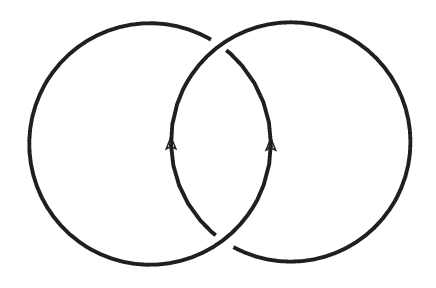}}
\vspace*{8pt}
\caption{Hopf link $L2a1$.\label{fig:Hopf-link}}
\end{figure}
{The knot group of the Hopy link $L2a1$, i.e. the fundamental group
$\pi_1(C(L2a1))$ of the Hopf link complement $C(L2a1)$, can be calculated to be} 
\[
\pi_{1}\left(C(L2a1)\right)=\langle a,b\,|\,aba^{-1}b^{-1}=[a,b]=e\rangle=\mathbb{Z}\oplus\mathbb{Z}
\]
In \cite{KirKla:90}, the representation of knot and link groups is discussed. 
As shown in \cite{BullockFrohmanKania-Bartoszynska1999}, there is a relation of this 
representations to the so-called skein modules. Skein modules can be seen as the deformation 
quantization of these representations. Then at the view of skein modules, the representation 
of the Hopf link can be interpreted as to put a $SU(2)$ representing on each component. 
That means that every component is related to a $SU(2)$ representation, 
i.e., the knot group $\pi_{1}(C(L2a1))$ is represented by $SU(2)\otimes SU(2)$.
From the superconducting
point of view, one puts one qubit $\psi_{1}$ on one ring and the
second qubit $\psi_{2}$ on the other ring. Again, the coupling is
realized by a Josephson junction between $\psi_{1},\psi_{2}$
described by the equation
\[
i\hbar\frac{d}{dt}\left(\begin{array}{c}
\psi_{1}\\
\psi_{2}
\end{array}\right)=\left(\begin{array}{cc}
\boldsymbol{E} & \boldsymbol{V}\\
\boldsymbol{V} & -\boldsymbol{E}
\end{array}\right)\left(\begin{array}{c}
\psi_{1}\\
\psi_{2}
\end{array}\right)
\]
where we tuned the energy levels so that $\psi_{1}$ has energy $E$
and $\psi_{2}$ has energy $-E$. Then the Hamiltonian is given by
\[
H_{2}=\boldsymbol{E}\cdot\sigma_{z}+\boldsymbol{V}\cdot\sigma_{x}
\]
acting on the 2-qubit state. But both states $\psi_{1},\psi_{2}$
are decomposed into $(|0\rangle,|1\rangle)$ vectors so that at the
level of these vectors we have the Hamiltonian 
$H_{1\otimes 2}=H_{2}\otimes H_{1}$ i.e.
\[
H_{1\otimes 2}=\left(\boldsymbol{E}\cdot\sigma_{z}+\boldsymbol{V}\cdot\sigma_{x}\right)\otimes\left(E\cdot\sigma_{z}+V\cdot\sigma_{x}\right)
\]
This Hamiltonian has the right structure, i.e. couplings like $\sigma_z \otimes\sigma_x$,
to realize 2-qubit operations. For example, the CNOT gate given
by the Hamiltonian $(1-\sigma_{z})\otimes(1-\sigma_{x})$. Finally
via the scheme above, one can realize a universal quantum computer
by linked superconductors coupled by Josephson junctions.\\
As noted above, there are three types of qubit, the phase, charge and flux qubit.
Now we start with the realization of the flux qubit and we made the following assumptions:
\begin{itemize}
\item every over-crossing or under-crossing is made into a Josephson junction,
i.e. the two superconductor parts are coupled;
\item the qubit is realized by a Flux qubit, abstractly described by the
Hamiltonian
\[
{\displaystyle H={\frac{q^{2}}{2C_{J}}}+\left({\frac{\Phi_{0}}{2\pi}}\right)^{2}{\frac{\phi^{2}}{2L}}-E_{J}\cos\left[\phi-\Phi{\frac{2\pi}{\Phi_{0}}}\right]},
\]
with $\Phi_{0}=\hbar c/2e,\Phi$ the flux quant and flux, $L$ inductance
of the ring, $C_{J}$ the junction capacity, $E_{J}$ junction energy
and $\phi$ phase shift.
\end{itemize}  
The states correspond to a symmetrical $|0\rangle$ and an anti-symmetrical
superposition $|1\rangle$ of zero or single trapped flux quanta,
sometimes denoted as clockwise and counterclockwise loop current states.
The different energy levels are given by different (integer) numbers of 
magnetic flux quanta trapped in the (knotted) superconducting ring. 
Therefore to realize the flux qubit, we have to understand the relation between
the shape of the knotted superconductor and the flux (or better the fluxion
quantization). At first we state that in the case of the trefoil knot, 
the flux of the knotted superconductor
can be controlled by the two loops $a,b$ (two generators of the knot group), 
the third loop $c$ is determined by the relation $c=a^{-1}ba$.\\
The flux properties of the knotted superconductor
was described in \cite{Kephart1985}. Main result of this work is the description
of these properties by considering the space around the knotted superconductor (by using
the Meissner effect).  
Then (from the formal point of view) we have to consider a function over the knot
complement $C(K)\to U(1)$ representing the vector potential which generates the flux. 
This map induces a map of the fundamental
groups $\pi_{1}(C(K))\to\pi_{1}(U(1))=\mathbb{Z}$. Here the mapping
into the integers is given by the integer part of the ratio $\Phi/\Phi_{0}$ 
(number of flux quanta). This map seem to imply that the flux or 
better the fluxion quantization does not depend on knot.
In \cite{Kephart1985},
the fluxion quantization was described for knotted superconductors,
which are knotted like the trefoil knot. Usually for a ring $C$, one has the
well-known relation (Stokes theorem)
\[
N=\frac{1}{\Phi_{0}}\ointop_{C=\partial S}A=\frac{1}{\Phi_{0}}\intop_{S}F
\]
between the electromagnetic potential (seen as 1-form) and the flux
through a surface $S$ (a disk $D^{2}$). This theorem can be generalized
to the knot $K$: for every knot there is a surface $S_{K}$ with
minimal genus $g$ so that $K=\partial S_{K}$. It is known that only
the unknot has a Seifert surface of genus $0$. The two knots $3_{1}$
(trefoil) and $4_{1}$(figure-8) have Seifert surfaces of genus $1$.
In case of the trefoil knot, a simple picture was found to express
the flux quantization, see \cite{Kephart1985}. There, the flux $\Phi_{SS}$
through the Seifert surface is decomposed into a linear combination
of two (approximately) conserved fluxes $\Phi_{R},\Phi_{Q}$ so that
$\Phi_{SS}=3\Phi_{R}+2\Phi_{Q}$. The coefficients are given by another
representation of the knot group 
$\pi_{1}(C(3_{1}))=\langle\alpha,\beta\,|\,\alpha^{3}=\beta^{2}\rangle$
(with different generators, see Fig. 2 in \cite{Kephart1985}). The
trefoil knot belongs to the class of torus knots, i.e. a closed curve
winding around the torus.\\ 
In case of non-torus knots like the figure-8 knot, we cannot use these ideas.
Instead we will follow another path. The knot complement has a boundary, which is
a torus. For the torus, we know the fluxion quantization. Fora general knot, we have to know
how the knot lies inside of the knot complement. In case of the flux qubit, we have to 
understand how the boundary torus (where we know the flux) lies inside of the 
knot group representations. Here, one has to use the so-called A-polynomial to 
get these information and to express the fluxion quantization. Let
$C(4_{1})$ be the knot complement of the figure-8 knot $4_{1}$.
As we remarked above this knot complement carries a hyperbolic structure
which will be used to define the A-polynomial. 
The hyperbolic structure is given by the choice of a homomorphism
$\pi_{1}(C(4_{1}))\to SL(2,\mathbb{C})$ (up to conjugation). Every
knot complement $C(K)$ has the boundary $\partial C(K)=T^{2}$. The
problem is now how the torus boundary lies inside of the space of
all hyperbolic structures (character variety) 
$ChV(K)=Hom(\pi_{1}(K),SL(2,\mathbb{C}))/SL(2,\mathbb{C})$.
Then the torus inside of $ChV(4_{1})$ is defined by the zero set
of a polynomial, the A-polynomial (for the details consult \cite{APolynomial1994,APolynomial1998}).
For the figure-8 knot $4_{1}$, the A-polynomial is given by
\[
A(M,L)=-2+M^{4}+M^{-4}-M^{2}-M^{-2}-L-L^{-1}
\]
and the decomposition of $A(\pm 1,L)=(L-1)^{2}L^{-1}$ gives the first
possible values $(2,-1)$ how the torus (via the slopes) lies in $ChV(4_{1})$
(interpreted as eigenvalues of the torus slopes). Now we can use these
eigenvalues to get the decomposition of the flux into the 
two (approximately) conserved fluxes $\Phi_{R},\Phi_{Q}$ so that
$\Phi_{SS}=2\Phi_{R}-\Phi_{Q}$ for the first possible values.
In contrast to the trefoil knot, there are more than one possible values
for the combination $\Phi_{SS}$.
In particular, this example showed that the fluxion quantization 
is more complex for hyperbolic knots (in contrast to torus knots). 
Then for the flux qubit one gets different combinations of the flux 
$\Phi_{SS}$ in dependence on the generator of the knot group.\\
The discussion above for the flux qubit showed that the relation between 
the flux, the operation and the qubit is complicated. Now we will discuss
another possibility to realize the operations. Above we discussed
the Hamiltonian operators $H_{1}$ and $H_{1\otimes 2}$ to describe the 
Josephson junctions. There, we found the interesting result that the Josephson
junction at the over or under crossing has the same effect then a gate  
(as element of the knot group representation) acting on the qubit.
The reason for this unexpected behavior is rooted in the energy/potential  
difference between the two states $|0\rangle$ and $|1\rangle$, respectively.
For a controlled behavior of the operations, one needs extra 
Josephson junctions. Every junction is located at the parts which
are represented by the generators $a,b,c$ of the knot group 
(see Fig. \ref{fig:three-generators-knot-group-trefoil}).  
As discussed above, we can trim the energy levels so that we get the Hamiltonian
\[
H_{1}=E\cdot \sigma_z+Ve^{i\phi}\sigma_x
\]
with a phase shift $\phi$ as induced by the coupling in the junctions.
Every junction gives rise to an operation $\exp(iH_{1}t)$ of one qubit which
is related to a representation of the knot group. The 2-qubit operations are 
induced by the linking where one has to add an Josephson junction near the linking 
(which is also a Josephson junction). This Josephson junction is described by the 
Hamiltonian $H_{1}$ and together with the linking we get the Hamiltonian $H_{1\otimes 2}$
above. Then we obtain the 2-qubit operation $\exp(iH_{1\otimes 2} t)$.\\
We will close this section with a remark about the decoherence. The knotting of 
the superconductor (via the Josephson junctions at the over and under crossings)
gives a self-coupling which has the ability to stabilize the state. We conjecture
that a qubit (flux or phase qubit) on knotted superconductor has an increased
decoherence time. {We will discuss it in a forthcoming work.}
}

\section{Conclusion}

In this paper we presented some ideas to use 3-manifolds for quantum
computing. As explained above, the best representative is the fundamental
group of a manifold. The fundamental group is the set of closed curves
up to deformation with concatenation as group operation up to homotopy.
It is known that every 3-manifold can be decomposed into simple pieces
so that every piece carries a geometric structure (out of 8 classes).
In principle, the pieces consist of complements of knots and links.
Then the fundamental group of the knot complement, known as knot group,
is an important invariant of the knot or link. Why not use this knot
group for quantum computing? In \cite{Planat2017,Planat2017a,Planat2018,Planat2019}
M. Planat et.al. studied the representation of knot groups and the
usage for quantum computing. Here we discussed a realization of knot
complements by knotted superconductors where the crossings are Josephson
junctions. The qubit is given by the flux qubit but we also discuss the phase qubit. 
As shown in \cite{AsselmeyerMaluga2021a},
the knot group determines the operations and we got all 1-qubit operations
for a knot. Then we discussed the construction of 2-qubit operations
by linking the two knotted superconductors. 


\section*{Acknowledgments}
I want to thank the anonymous referee for the helpful remarks which increases the readability of this paper.
%


\end{document}